\documentclass[10pt,journal,compsoc]{IEEEtran}

\ifCLASSOPTIONcompsoc
  \usepackage[nocompress]{cite}
\else
  \usepackage{cite}
\fi

\ifCLASSINFOpdf
  \usepackage[pdftex]{graphicx}
\else
\fi

\usepackage{amsmath,amsfonts}

\ifCLASSOPTIONcompsoc
  \usepackage[caption=false,font=footnotesize,labelfont=sf,textfont=sf]{subfig}
\else
  \usepackage[caption=false,font=footnotesize]{subfig}  
\fi
\usepackage{capt-of}

\usepackage{url}

\usepackage[boxed,vlined,ruled]{algorithm2e}
\SetAlCapSkip{0.5em}

\SetCommentSty{mycommfont}

\usepackage{stmaryrd}
\usepackage{enumitem}
\usepackage{booktabs}
\usepackage{xspace}

\DeclareMathOperator*{\argmax}{arg\,max}
\DeclareMathOperator*{\exttsp}{ExtTSP}
\DeclareMathOperator*{\tsp}{TSP}
\DeclareMathOperator*{\size}{blocks}
\DeclareMathOperator*{\len}{len}
\DeclareMathOperator*{\degree}{degree}

\newcommand{\Oh}{{\ensuremath{\mathcal{O}}}}
\newcommand{\NP}[0]{\texttt{NP}\xspace}
\newcommand{\alg}[1]{{\bf \texttt{#1}}\xspace}
\newcommand{\prob}[1]{\textsc{#1}\xspace}
\newcommand{\TSP}{\textsc{TSP}\xspace}
\newcommand{\ExtTSP}{\textsc{ExtTSP}\xspace}

\newcommand{\sra}{{\ensuremath{\shortrightarrow}}}

\begin{document}
\title{Improved Basic Block Reordering}
%
%
%
%

\author{Andy~Newell
        and~Sergey Pupyrev
\IEEEcompsocitemizethanks{
\IEEEcompsocthanksitem A. Newell is with Facebook, Inc., Menlo Park, CA, USA.
\protect\\
E-mail: newella@fb.com
\IEEEcompsocthanksitem S. Pupyrev is with Facebook, Inc., Menlo Park, CA, USA.
\protect\\
E-mail: spupyrev@fb.com}%
}

\makeatletter
\def\ps@IEEEtitlepagestyle{
	\def\@oddfoot{\mycopyrightnotice}
	\def\@evenfoot{}
}
\def\mycopyrightnotice{
	{\footnotesize
		\begin{minipage}{\textwidth}
			\vspace{0.5cm}
\copyright~2020 IEEE.  Personal use of this material is permitted.  Permission from IEEE must be obtained for all other uses, in any current or future media, including reprinting/republishing this material for advertising or promotional purposes, creating new collective works, for resale or redistribution to servers or lists, or reuse of any copyrighted component of this work in other works.			
		\end{minipage}
	}
}

\IEEEtitleabstractindextext{%
\begin{abstract}
	Basic block reordering is an important step for profile-guided binary optimization.
	The state-of-the-art goal for basic block reordering is to maximize the number of fall-through branches.
	However, we demonstrate that such orderings may impose suboptimal performance on instruction and I-TLB caches.
	We propose a new algorithm that relies on a model combining the effects of fall-through and caching behavior.
	As details of modern processor caching is quite complex and often unknown, we show how to use machine learning in 
	selecting parameters that best trade off different caching effects to maximize binary performance.
	
	An extensive evaluation on a variety of applications, including Facebook production workloads, the open-source 
	compilers Clang and GCC, and SPEC CPU benchmarks, indicate that the new method outperforms existing block reordering 
	techniques, improving the resulting performance of applications with large code size.
	We have open sourced the code of the new algorithm as a part of a post-link binary optimization tool, BOLT.
\end{abstract}

}

\maketitle

\IEEEdisplaynontitleabstractindextext

%
\IEEEpeerreviewmaketitle

\IEEEraisesectionheading{\section{Introduction}\label{sec:introduction}}

\IEEEPARstart{P}{rofile}-guided binary optimization (PGO) is an important step for improving performance of large-scale applications that tend to contain huge amounts of code. 
Such techniques, also known as feedback-driven optimization (FDO), are designed to improve code locality which leads to better utilization of CPU instruction caches.
In practice tools like AutoFDO~\cite{DXT16}, Ispike~\cite{LMPCL04}, PLTO~\cite{SDAL01}, HFSort~\cite{OM17}, and BOLT~\cite{PANO18} speed up binaries by $5\%-15\%$ depending on workload and CPU architecture, and thus, are widely used for a variety of complex applications.

PGO is comprised of a number of optimization passes such as function and basic block reordering, identical code folding, function inlining, unreachable code elimination, register allocation, and others. 
Typical targets for optimizations are an instruction cache (I-cache) used to hold executable instructions and a translation lookaside buffer (I-TLB) used to speed up virtual-to-physical address translation for instructions. 
The reordering passes directly optimize code layout, and thus impact performance the most~\cite{LMPCL04,PANO18}.
Therefore, even small improvements in the underlying algorithms for code reordering significantly affect the benefit of PGO tools.


Current techniques for basic block reordering optimize a specific dimension of CPU performance such as
(i)~cache line utilization by increasing the average number of instructions executed per cache line, 
(ii)~the branch predictor by reducing the number of mispredicted branches, and 
(iii)~the instruction cache miss rate by minimizing cache line conflicts.
An application's overall performance depends on a combination of these dimensions. 
Since modern processors employ a complex and often non-disclosed strategy for execution, it is challenging to consider all of these effects at once when optimizing an ordering of basic blocks.
\emph{In this paper, we make the first, to the best of our knowledge, attempt to design and implement a block reordering algorithm that directly optimizes the performance of an application}.

Our approach consists of two main steps. 
Firstly, we learn a proxy metric that describes the relationship between the performance of a binary and the ordering of its basic blocks. 
This is achieved by 
(i)~identifying a set of features representing how basic block ordering can influence performance, 
(ii)~collecting training data by running extensive experiments and measuring the performance, and
(iii)~using machine learning to select the best combination of the features for a score that best predicts CPU performance.
Secondly, we suggest an efficient algorithm that, given a control flow graph for a procedure, builds an improved ordering of the basic blocks optimizing the learned metric. 
Since the constructed metric correlates highly with the performance of a binary, we observe overall efficiency gains, despite possible regressions of individual CPU characteristics. 

The contributions of the paper are the following.

\begin{itemize}[leftmargin=6mm]
	\item We identify an opportunity for improvement over the classical approach for basic block reordering, initiated by Pettis and Hansen~\cite{PH90}. 
	Then we extend the model and suggest a new optimization problem with the objective closely related to the performance of a binary.
	
	\item We then develop a new practical algorithm for basic block reordering.
	The algorithm relies on a greedy technique for solving the optimization problem. 
	We describe the details of our implementation, which scales to real-world instances without significant impact on the running time of a binary optimization tool.
	
	\item We propose a Mixed Integer Programming formulation
	for the aforementioned optimization problem, which is capable of finding optimal solutions on small functions.
	Our experiments with the exact method demonstrate that the new suggested heuristic finds an
	optimal ordering of basic blocks in $98\%$ of real-world functions with $30$ or fewer blocks.
	
	\item Finally, we extensively evaluate the new algorithm on a variety of applications, 
	including Facebook production workloads, open-source compilers, Clang and GCC, and SPEC CPU 2017
	benchmarks. The experiments	indicate that the new method outperforms the state-of-the-art 
	block reordering techniques, improving the resulting performance by $0.5\%-1\%$.
	We have open sourced the code of our new algorithm as a part of BOLT~\cite{PANO18,BOLTSrc}.
\end{itemize}	

The paper is organized as follows. We first discuss limitations of the existing model
for basic blocks reordering and suggest an improvement in Section~\ref{sect:ml}.
We describe an efficient heuristic (Section~\ref{sect:algo}) and an exact algorithm (Section~\ref{sect:mip}) 
for solving the new problem. Next, in Section~\ref{sect:exp}, we present experimental results, which
are followed by a discussion of related work in Section~\ref{sect:related}.
We conclude the paper and discuss possible future directions in Section~\ref{sect:conclude}.

\section{Learning an Optimization Model}
\label{sect:ml}

The state-of-the-art approach for basic block reordering is based on the idea of collocating frequently executed blocks together.
The goal is to position blocks so that the hottest successor of a block will most likely be a fall-through branch, that is, located 
right next to the predecessor. This strategy reduces the number of taken branches and the working set size of the I-cache, while 
relieving pressure from the branch predictor unit. More formally, the reordering problem can be formulated as follows. Given a 
directed control flow graph comprising of basic blocks and frequencies of jumps between the blocks, find an ordering of the blocks 
such that the number of fall-through jumps is maximized. This is the maximum directed \prob{Traveling Salesman Problem} (\TSP), a widely studied \NP-hard combinatorial optimization problem. 

The simplicity of the model and solid practical results made \TSP-based algorithms very popular in the code optimization community. 
To the best of our knowledge, Boesch and Gimpel~\cite{BG77} are the first ones to 
formulate the problem of finding an ordering of basic block as the
path covering problem on a control flow graph, which is equivalent to solving \TSP. They describe an
optimal algorithm on acyclic directed graphs and suggest a heuristic for general digraphs.
Later the same path covering model has been studied in a series of papers suggesting optimal algorithms
for special classes of digraphs and heuristics for general digraphs~\cite{YJSK97,HMK79}.
In their seminal paper from 1990~\cite{PH90}, Pettis and Hansen present two heuristics 
for positioning of basic blocks. We observe that both heuristics are designed to solve (possibly non-optimally) 
\TSP. Later, one of the heuristics (seemingly producing better results) has been extended 
by Calder and Grunwald~\cite{CG94}, Torrellas et al.~\cite{TXD98}, and Luk et al.~\cite{LMPCL04}.
We stress that the majority of existing algorithms for block reordering utilize the \TSP model.
A variant of the Pettis-Hansen algorithm is used by most of the modern binary
optimizers, including PLTO~\cite{SDAL01}, Ispike~\cite{LMPCL04}, BOLT~\cite{PANO18}, and the
link-time optimizer (LTO) of the GCC compiler~\cite{RLN14}.

\begin{figure}[!t]
	\centering
	\subfloat{
		\includegraphics[width=0.49\columnwidth,page=3]{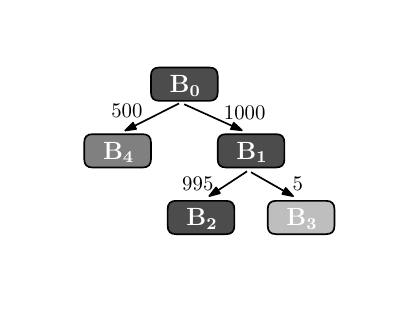}
	}
	\subfloat{
		\includegraphics[width=0.49\columnwidth,page=4]{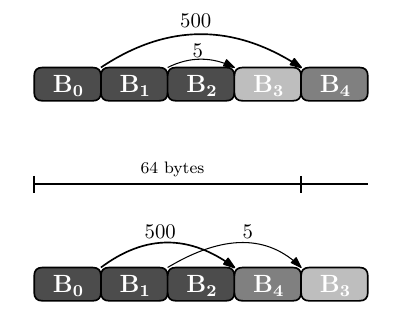}
	}
	\caption{Two orderings of basic blocks with the same \TSP score ($1995$) resulting in different I-cache utilization.
		All blocks have the same size of $16$ bytes and colored according to their hotness in the profile.}
	\label{fig:tsp}
\end{figure}

\begin{figure}[!t]
	\centering
	\subfloat{
		\includegraphics[width=0.34\columnwidth,page=1]{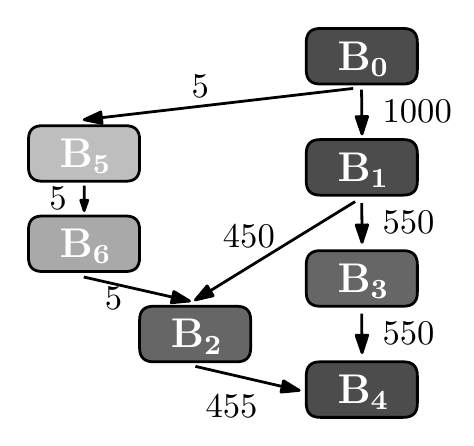}
	}	
	\subfloat{
		\includegraphics[width=0.64\columnwidth,page=2]{pics/tsp_vs_exttsp}
	}	
	\caption{A control flow graph with jump frequencies (left) and 
		two possible orderings of basic blocks (right).
		All blocks have the same size (in bytes) and colored according to their hotness in the profile.
		An optimal \TSP-based layout (upper right) utilizes three cache lines for the hot code, while
		an arguably better layout (lower right) can be built with a new \ExtTSP model.
	}
	\label{fig:tsp_vs_exttsp}
\end{figure}

Notice that solving \TSP alone is not sufficient for constructing a good ordering of basic blocks.
It is easy to find examples of control flow graphs with multiple different orderings that are all optimal with
respect to the \TSP objective. Consider for example a control flow graph in Fig.~\ref{fig:tsp} in which
the maximum number of fall-through branches is achieved with two orderings that
utilize a different number of I-cache lines in a typical execution.
For these cases, an algorithm needs to take into consideration
non-fall-through branches to choose the best ordering. However, maximizing the number of 
fall-through jumps is not always preferred from the performance point of view. Consider a control flow
graph with seven basic blocks in Fig.~\ref{fig:tsp_vs_exttsp}.
It is not hard to verify that the ordering with the maximum number
of fall-through branches is one containing two concatenated chains, $B_0 \sra B_1 \sra B_3 \sra B_4$
and $B_5 \sra B_6 \sra B_2$ (upper-right in Fig.~\ref{fig:tsp_vs_exttsp}). 
Observe that for this placement, the hot part of the function occupies three 64-byte cache lines. Arguably a better ordering
is the lower-right in Fig.~\ref{fig:tsp_vs_exttsp}, which uses only two cache lines for the five hot blocks, 
$B_0, B_1, B_2, B_3, B_4$, at the cost of breaking the lightly weighted branch $B_6 \sra B_2$.

How do we identify the best ordering of basic blocks? The question is fairly difficult and even
experts may have hard time determining which ordering leads to the maximum performance of a binary. A naive approach
is to exhaustively evaluate every valid block placement and then profile the binary to collect relevant performance
metrics. Obviously due to the enormous search space, this approach is infeasible for practical use. A natural
improvement is to reduce the search space and experiment only with the most promising orderings. This technique, also
known as iterative compilation or autotuning, is a natural task for machine learning~\cite{AKCPS18,WB18}.
While in certain scenarios the overhead is justifiable, we found this approach impractical for our production systems
due to long build and deployment times.
Therefore, we use another strategy for optimization by developing a score function that is used
as a proxy for estimating the quality of an ordering. The idea is to perform extensive experiments profiling
an application in order to understand what aspects and features of the block placement affect the resulting performance. After
this first phase, employ a machine learning technique to build an optimization model and derive a quality 
metric for an ordering. As a final step, design an algorithm to optimize the constructed score function.

Next we describe the process in detail, explaining the data collection phase and
presenting the developed score function. 
Since our new approach for basic block reordering is implemented in a post-link optimizer BOLT~\cite{PANO18} and evaluated on modern
Intel x86 processors, we describe the 
steps that are typical for this setup. Note however that the
approach is not tied to the binary optimizer and can be similarly applied in other environments.

\subsection{Data Collection}

Following most of the recent works on profile-guided code optimizations~\cite{LMPCL04,DXT16,OM17,PANO18},
we rely on sampling techniques for collecting profile data. Although the sampling-based approach is
typically less accurate than the instrumentation-based one, it incurs significantly less memory and
performance overheads, making it the preferred way of profiling binaries in actual production
environments. We utilize hardware support of Intel x86 processors to collect Last Branch Records (LBR), which
is a list of the last 16 taken branches. From the list of branches, that are sampled according to a specified
event, we infer the frequencies of jumps between basic blocks. Specifically, we extract a weighted
directed control flow graph for every function in the profiled binary. The vertices (basic blocks) and the
edges (branches) of the graph, along with the sizes of the blocks, are extracted 
statically via the BOLT infrastructure, which is based on LLVM~\cite{LA04}. The weights between the basic blocks
correspond to the total number of times the jumps appear in collected LBRs.
We stress that before processing, we augment collected LBRs with fall-through jumps, as LBRs only contain
information about taken branches; to this end, we utilize a simple algorithm similar to one described
by Chen~et~al.~\cite{Chen13}.
Notice that we ignore indirect branches, procedure calls, and returns while constructing the control flow graph.

We have experimented with several different events to collect LBRs, including \emph{cycles}, \emph{retired instructions}, and
\emph{taken branches}, and using various levels of
precise event based sampling. We observed that independently of the utilized event and processor microarchitecture, the
extracted jump frequencies do not always follow the expected distribution;
refer to \cite{Chen13,NYMZ15,PANO18} for concrete examples and possible explanations of the phenomenon.
Hence, we use by default the \emph{cycles} event to sample LBRs.

\begin{figure}[!t]
	\centering
	\subfloat[]{
		\includegraphics[width=0.45\columnwidth,page=1]{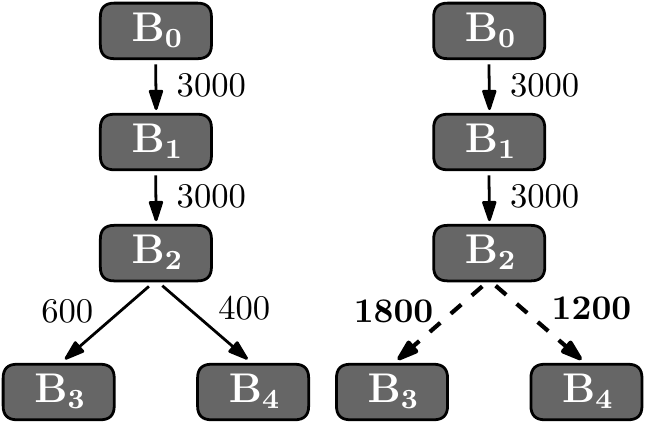}
		\label{fig:mcfa}
	}	
\hfill
	\subfloat[]{
		\includegraphics[width=0.45\columnwidth,page=2]{pics/mcf}
		\label{fig:mcfb}
	}	
	\caption{Two examples of an original incomplete profile (left) and its realistic correction (right) that
		cannot be reconstructed using the \prob{Minimum Cost Maximum Flow} model. Bold numbers show most
		realistic adjustments of the edge weights satisfying flow conservation constraints.}
	\label{fig:mcf}
\end{figure}

A common technique for ensuring edge weights are more realistic is solving the \prob{Minimum Cost
Maximum Flow} problem on the control flow graph, which was initiated by Levin et al.~\cite{LNH08}
and later adopted by several groups~\cite{Chen13,Nov14,NYMZ15,LPZ16}. In contrast with the earlier works, 
our experiments with the flow-based approach did not produce performance gains in comparison with
the original (possibly biased) data. A problematic example for the approach is illustrated in Fig.~\ref{fig:mcfa}, 
where the arguably most realistic adjustment is highlighted. Depending on how the costs of the edges
are assigned, an algorithm for \prob{Minimum Cost Maximum Flow} will either produce
frequencies $2600$ and $400$ or $600$ and $2400$ for jumps $B_2 \sra B_3$ and $B_2 \sra B_4$, respectively.
However, it is desirable to keep the probabilities of the branches at $B_2$ (approximately) the same.
A related issue is shown in Fig.~\ref{fig:mcfb}. Here an algorithm may decide to send some flow along edge
$B_2 \sra B_4$, thus making basic block $B_4$ hot. This adjustment prohibits future compiler optimizations that 
position hot and cold parts of the function in different sections of the binary.
Therefore, we avoid modifying jump frequencies
via the flow-based approach, leaving for the future the task of increasing profile precision. 

\subsection{Engineering a Score Function}
\begin{figure}[!t]
	\centering
	\begin{minipage}{.46\columnwidth}
		\centering
		\includegraphics[width=\columnwidth,page=1]{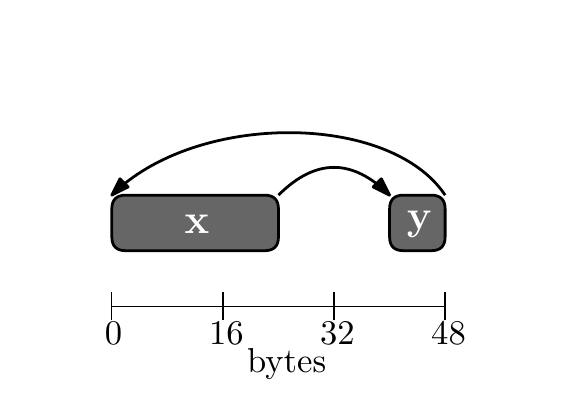}
		\captionof{figure}{The lengths of a forward jump, $x \sra y$, and
			a backward jump, $y \sra x$, are $16$ and $48$ bytes, respectively.}
		\label{fig:featuresa}
	\end{minipage}%
\hfill
	\begin{minipage}{.45\columnwidth}
		\centering
		\includegraphics[width=\columnwidth,page=2]{pics/features}
		\captionof{figure}{The dependency between the length of a jump and its importance for the \ExtTSP model.}
		\label{fig:featuresb}
	\end{minipage}
\end{figure}

Our goal is to design a function $x \rightarrow f(x)$, that takes in a feature vector
$x$, characterizing an ordering of basic blocks, and produces a real value $f(x)$, indicating 
an expected performance of a binary for the ordering. We assume that the execution time
of a single basic block is independent of the block ordering within a function. Thus, the
ordering only affects branches between the blocks, which may incur some delay
in the execution, for example due to a miss in the instruction cache.
However, not all branches equally affect the performance. An important feature of a branch is the \emph{jump length}, 
that is, the distance (in bytes) between the end of the source block to the beginning of the 
target block; see Fig.~\ref{fig:featuresa}.
For example, it is a common belief that
zero-length jumps (equivalently, fall-through branches) impose the smallest performance overhead. This is the 
main motivation for the \TSP model, whose objective can be formally expressed as follows:
$$
\tsp = \sum_{(s, t)} 
w(s, t) \times
\begin{cases}
1 & \text{if } \len(s, t) = 0,\\
0 & \text{if } \len(s, t) > 0,\\
\end{cases}
$$
where $w(s, t)$ is the frequency and $\len(s, t)$ is the
length of branch $s \sra t$.
An optimal ordering corresponds to the \emph{maximum} value
of the expression; thus, we call it the \emph{score} of \TSP.
The performance, however, might also depend on other characteristics of a branch, which we discuss next. 
In our study, we consider the following features.

\begin{itemize}[leftmargin=6mm] 
	\item The length of a jump impacts the performance of instruction caches. Longer jumps
	are more likely to result in a cache miss than shorter ones. In particular, a jump with the length
	shorter than $64$ bytes has a chance to remain within the same cache line.
	
	\item The direction of a branch plays a role for branch predicting. 
	A branch $s \sra t$ is called \emph{forward} if $s < t$, that is, block $s$ precedes block $t$ in
	the ordering; otherwise, the branch is called \emph{backward}.
	
	
	\item The branches can be classified into \emph{unconditional} (if the out-degree is one) and \emph{conditional}
	(if the out-degree is two). A special kind of branches is between consecutive blocks in the ordering
	that are called \emph{fall-through}; in this case, 
	a jump instruction is not needed.	
\end{itemize}	

We introduce a new score that estimates the quality of a basic block ordering taking into account the
branch characteristics. In the most generic form, the new function, called \prob{Extended TSP} (\ExtTSP),
is expressed as follows:

$$
\exttsp = \sum_{(s, t)} 
w(s, t) \times K_{s, t} \times h_{s, t}\big(\len(s, t)\big), 
$$
where the sum is taken over all branches in the control flow graph.
Here $w(s, t)$ is the frequency of branch $s \sra t$ and $0 \le K_{s, t} \le 1$ is a weight coefficient
modeling the relative importance of the branch for optimization. We distinguish six types of branches
arising in code: conditional and unconditional versions of fall-through, forward, and backward branches. Thus, we introduce
six coefficients for \ExtTSP.
The lengths of the jumps are accounted in the last term of the expression, which
increases the importance of short jumps. A non-negative function
$h_{s, t}\big(\len(s, t)\big)$
is defined by value of $1$ for zero-length jumps, value of $0$ for jumps exceeding a prescribed
length, and it monotonically decreases between the two values.
To be consistent with the objective of \TSP, the \ExtTSP score needs to be \emph{maximized} for the best performance.
Notice that \ExtTSP is a generalization of \TSP, as the latter can be modeled by setting $K_{s,t}=1, h\big(\len(s, t)\big)=1$ for
fall-through branches and $K_{s,t}=0$ otherwise.

In general we cannot manually select the most appropriate constants of \ExtTSP that best model
the performance of modern processors. Next we describe a process for
learning these constant values that lead to the best performance.

\begin{figure*}[!t]
	\centering
	\subfloat[\textsf{Clang}]{
		\includegraphics[width=0.95\columnwidth]{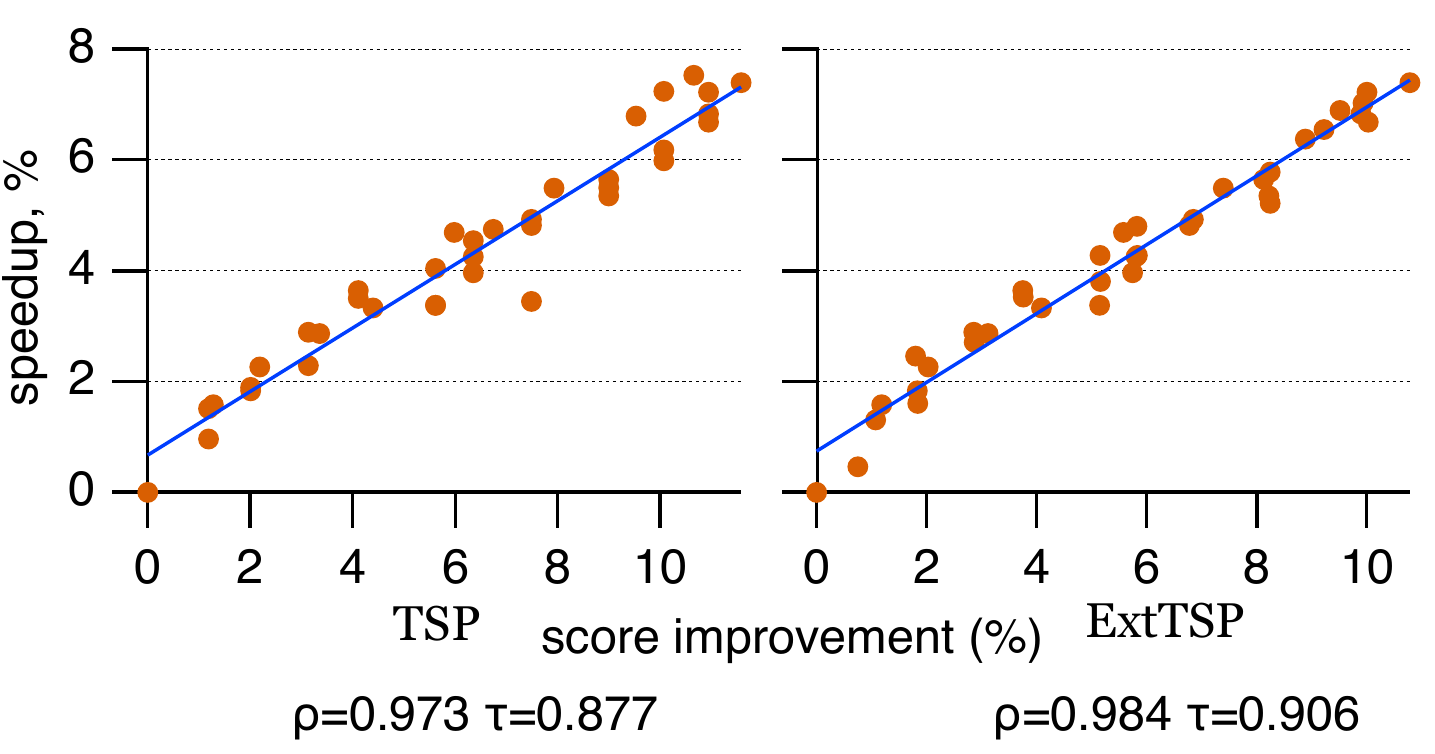}
	}
	\hfill
	\subfloat[\textsf{HHVM}]{
		\includegraphics[width=0.95\columnwidth]{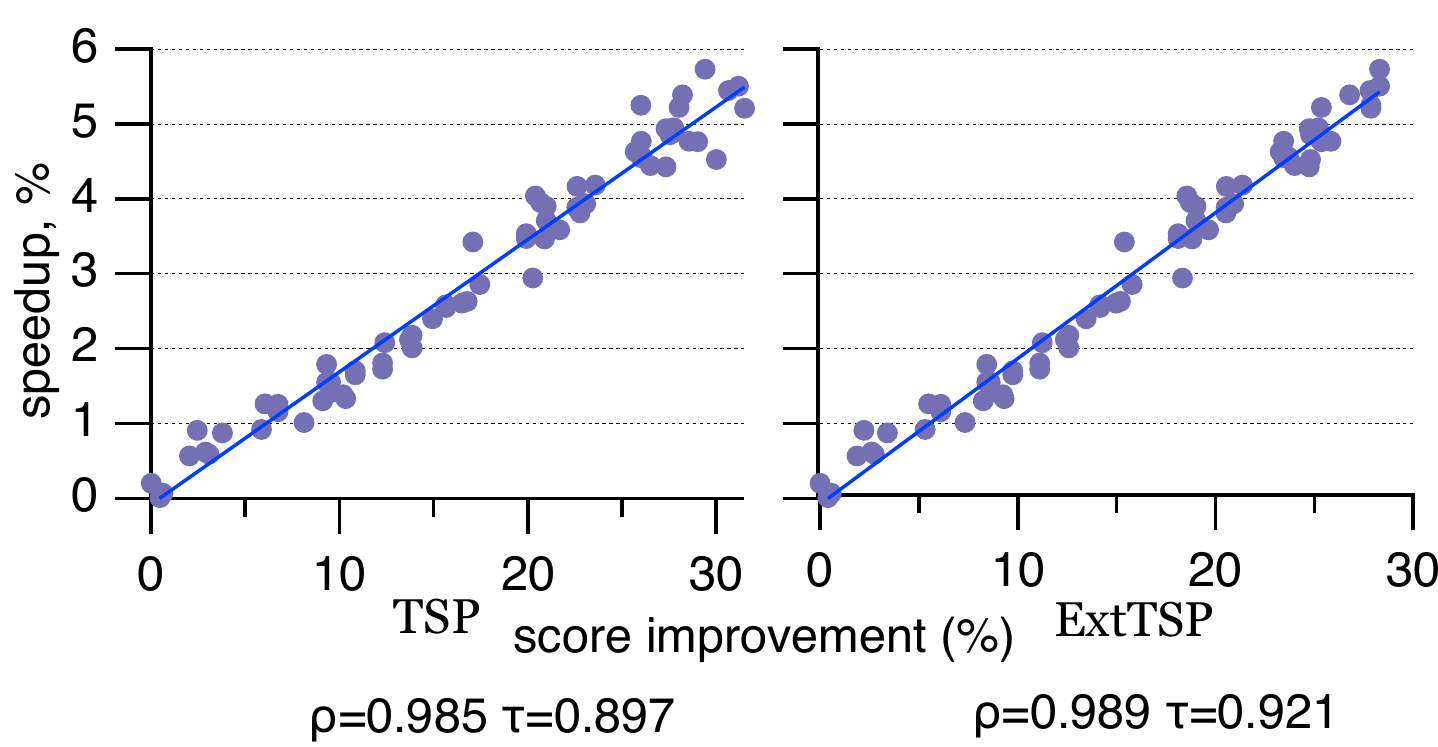}
	}	
	\caption{The relationship between the performance (instructions per cycle) of two binaries and
		the \TSP and \ExtTSP scores measured for various orderings of basic blocks. The
		values correspond to the relative improvements over the 
		non-optimized binary with the original ordering provided by the compiler.}
	\label{fig:cpu_vs_tsp}
\end{figure*}

\subsection{Learning Parameters}
\label{sect:bb}

As a preliminary step of our study, we run multiple experiments with two binaries, the \textsf{Clang} compiler and
the HipHop Virtual Machine (\textsf{HHVM})~\cite{AEM14}. Each experiment consists of constructing a distinct
ordering of basic blocks, running a binary, and measuring its performance metrics via the 
Linux \texttt{perf} tool. In order to build a variety of block orderings for the same binary, we
utilize five algorithms (described in Section~\ref{sect:techniques}) and apply them for a certain percentage 
of randomly selected functions. In total, we evaluated $50$ distinct block orderings and
conducted $250$ experiments (five per ordering) for each of the two binaries. 

Our first finding is that the traditional \TSP score has a relatively high correlation with
the performance of the binaries; see Fig.~\ref{fig:cpu_vs_tsp}. However, there are several
unexpected outliers that cannot be explained by the model.
In order to choose suitable parameters for the \ExtTSP score, we employ the so-called black-box 
solver developed by Letham et al.~\cite{LKOB18}, which is a powerful tool for optimizing functions
with computationally expensive evaluations. Formally, our problem can be stated as finding  
parameters for \ExtTSP that have the highest correlation with the performance of a binary
in the experiments. Here we try to maximize the Kendall rank correlation coefficient, $\tau$, between the
observed performance (instructions per cycle) and the predicted improvement given by the \ExtTSP score.
Notice that the Pearson correlation coefficient, $\rho$, is not the best choice for optimization, 
as the relationship between observed and predicted values might not be linear.
The black-box solver, which is based on Bayesian optimization, is able to compute values
for a collection of continuous parameters that maximize the correlation coefficient. This is
done via a careful exploration of the search space taking into account noise in real-world
experiment outcomes.
In our study we introduce six variables for weight coefficients, $K$, of $\exttsp$.
The jump-length function, $h(\cdot)$, is considered to be of the form
$\left(1 - \left(\frac{\len(jump)}{M}\right)^\alpha \right)$ with two
variables $M > 0$ and $\alpha > 0$ for different types of branches.

The black-box solver found a model that
better predicts the observed values; see Fig.~\ref{fig:cpu_vs_tsp}. The new model
increases the Kendall correlation coefficient $\tau$ from $0.877$ to $0.906$ for \textsf{Clang}
and from $0.897$ to $0.921$ for \textsf{HHVM}.
The models constructed for \textsf{Clang} and \textsf{HHVM} are not identical, though they
share many similarities. 
Next we present a unified variant of the model in which we round and combine parameters 
with similar weights (having difference less than $0.05$), and
exclude ones having small values (less than $0.05$).
We did not notice a discrepancy between the actual and the rounded \ExtTSP models, 
meaning that the resulting solution works well for both of the binaries and
is robust to the choice of constants.
Recall that better block orderings correspond to higher values of \ExtTSP. 

{
\small
$$
\!\!\!
\exttsp {=} \! \sum_{(s, t)} 
\!
w(s, t) {\times}
\begin{cases}
1 & \mbox{\small $\!\!\!\text{if } \len(s, t) = 0$}, \\
0.1 \! \cdot \! \left(1{-}\frac{\len(s, t)}{1024} \right) & \mbox{\small $\!\!\!\text{if } 0 {<} \len(s, t) {\le} 1024$} \\
& \mbox{\small $\!\!\! \text{and } s < t$},\\
0.1 \! \cdot \! \left(1{-}\frac{\len(s, t)}{640} \right) & \mbox{\small $\!\!\!\text{if } 0 {<} \len(s, t) {\le} 640$} \\
& \mbox{\small $\!\!\! \text{and } t < s$}, \\
0 & \mbox{\small \!\!\!\text{otherwise.}} \\
\end{cases}
$$
}

Intuitively, \ExtTSP resembles the traditional \TSP model, as the number
of fall-through branches is the dominant factor. The main difference is that \ExtTSP rewards longer jumps. 
The impact of such jumps is significantly
lower and it linearly decreases with the length of a jump.
Next we summarize our high-level observations regarding the new score function.

\begin{itemize}[leftmargin=6mm]
	\item The suggested parameters for \ExtTSP correlate well with the overall performance of a binary
	in a production-like environment, though we also observe moderate correlation (in the order of $\rho = 0.8$) 
	between the values of \ExtTSP and the measured number of I-cache misses. 
	
	\item We have not observed significant differences between the importance of conditional and unconditional branches that
	seem to be similarly relevant for the quality of a block ordering. It contradicts to the intuition
	of Calder and Grunwald~\cite{CG94} who assign noticeably different weights depending on the type of a
	branch.
	
	\item The maximum length of a jump affecting \ExtTSP is fairly large: $16$ and $10$ cache lines
	for forward and backward branches, respectively. The importance of a non-fall-through branch linearly 
	decreases with its length; see Fig.~\ref{fig:featuresb}.
	We experimented with non-linear decreasing functions but did not discover a significant
	improvement; hence, we use the simpler variant.
	
	\item We found that forward branches are more important for
	an ordering than backward ones; see Fig.~\ref{fig:featuresb} for a dependency of the weights of the two
	types of branches on the \ExtTSP score. Both types of non-fall-through branches are noticeably less
	important than fall-throughs in the constructed model, which is reflected in the low coefficient ($0.1$) in
	the expression. 
	
\end{itemize}

Finding an optimal solution for the \prob{Extended TSP} problem is \NP-hard. Next we describe our heuristic.

\section{A Heuristic for \ExtTSP}
\label{sect:algo}

Our algorithm finds an optimized ordering of basic blocks for every
function in the binary. It operates with a weighted control flow graph
$G=(V, E, w)$ containing a set of basic blocks, $V$, and directed edges, $E$,
representing branches between the blocks. An edge $(s, t) \in E$ 
corresponds to a jump from a block $s \in V$ to a block $t \in V$
and its weight, $w(s, t)$, corresponds to the frequency of the jump.
We assume that the sizes (in bytes) of the basic blocks are a part
of the input. The goal of the algorithm is to find an ordering of $V$
with an improved $\exttsp$ score (as defined in Section~\ref{sect:ml})
while keeping a given entry point, $v^* \in V$, the first in the ordering.

\begingroup
\setlength{\textfloatsep}{10pt}
\DecMargin{1.1em}
\begin{algorithm}[t]
	\caption{Basic Block Reordering}
	\label{algo:order}
	
	\SetKwInOut{Input}{Input}
	\SetKwInOut{Output}{Output}
	\SetKwProg{Fn}{Function}{}{end}
	\SetKwFunction{ComputeMergeGain}{ComputeMergeGain}
	\SetKwFunction{ReorderBasicBlocks}{ReorderBasicBlocks}
	\SetKwFunction{Merge}{Merge}
	
	\Input{control flow graph $G=(V, E, w)$,\\ the entry point $v^* \in V$}
	\Output{ordering of basic blocks $(v^*=B_1, B_2, \dots, B_{|V|})$}
	\BlankLine
	
	\Fn(){\ReorderBasicBlocks}{
		\For(\tcc*[f]{initial chain creation}){$v \in V$}{
			$Chains \leftarrow Chains \cup (v)$\;
		}
		\While(\tcc*[f]{chain merging}){$|Chains| > 1$}{
			\For{$c_i, c_j \in Chains$}{
				$gain[c_i, c_j] \leftarrow \ComputeMergeGain(c_i, c_j)$\;
			}
			\BlankLine
			\tcc{find best pair of chains}
			$src, dst \leftarrow \underset{i,j}{\argmax}~gain[c_i, c_j]$\;
			
			\tcc{merge the pair and update chains}
			$Chains \leftarrow Chains \cup \Merge(src, dst) \setminus \{src, dst\}$\;
		}
		\Return{ordering given by the remaining chain\;}
	}
	
	\BlankLine
	\BlankLine
	\BlankLine
	\Fn(){\ComputeMergeGain{$src, dst$}}{
		\tcc{try all ways to split chain $src$}
		\For{$i=1$ \emph{\KwTo} $\size(src)$}{
			\tcc{break the chain at index $i$}
			$s_1 \leftarrow src[1:i]$\;
			$s_2 \leftarrow src[i+1:\size(src)]$\;
			\tcc{try all valid ways to concatenate}
			$score_i \!\! \shortleftarrow \! \max \!
			\begin{cases}
			\exttsp(s_1, s_2, dst) \; \text{if} \enskip v^* \! \not \in\! dst\\
			\exttsp(s_1, dst, s_2) \; \text{if} \enskip v^* \! \not \in\! dst\\
			\exttsp(s_2, s_1, dst) \; \text{if} \enskip v^* \! \not \in\! s_1,dst\\
			\exttsp(s_2, dst, s_1) \; \text{if} \enskip v^* \! \not \in\! s_1,dst\\
			\exttsp(dst, s_1, s_2) \; \text{if} \enskip v^* \! \not \in\! src\\
			\exttsp(dst, s_2, s_1) \; \text{if} \enskip v^* \! \not \in\! src\\
			\end{cases}
			$
		}
		
		\BlankLine
		\tcc{the gain of merging chains $src$ and $dst$}
		\Return{$\underset{i}{\max}~score_i - \exttsp(src) - \exttsp(dst)$\;}
	}
\end{algorithm}
\IncMargin{1.1em}
\endgroup

On a high level, our algorithm is a greedy heuristic that works with chains
(ordered sequences) of basic blocks; see Algorithm~\ref{algo:order} for an
overview. Initially all chains are isolated basic blocks. Then we iteratively 
merge pairs of chains so as to improve the $\exttsp$ score.
On every iteration, we pick a pair of chains whose merging yields the biggest 
increase in $\exttsp$, and the pair is merged into a new chain. The procedure
stops when there is only one chain left, which determines the resulting
ordering of basic blocks.

\begin{figure}[!h]
	\centering
	\includegraphics[height=2.4cm,page=5]{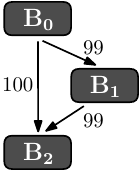}
	\caption{An example of a control flow graph in which naive chain concatenation produces suboptimal ordering.}
	\label{fig:greedy_fail}
\end{figure}

An important aspect of our approach is the way two chains
are merged; see function \textsc{ComputeMergeGain} of Algorithm~\ref{algo:order}.
In order to merge a pair of chains, $src$ and $dst$, we first split
chain $src$ into two subchains, $s_1$ and $s_2$, that retain the ordering 
of blocks given by $src$. Then we consider all six possible ways of combining the 
three chains, $s_1$, $s_2$, and $dst$, into a single one, discarding the ones
that do not place entry point $v^*$ at the beginning.
A chain with the largest $\exttsp$ over all possible splitting indices of $src$ 
and permutations of $s_1$, $s_2$, $dst$ is chosen as the result. 
The motivation here is to increase the search space 
in comparison to the simpler concatenation of two chains.
The simplest example in which chain splitting helps is depicted in Fig.~\ref{fig:greedy_fail}.
A greedy concatenation merges block $B_0$ with $B_2$ on the first iteration, which
results in the final ordering $(B_0,B_2,B_1)$.
In contrast, chain splitting allows to build an ordering $(B_0,B_1,B_2)$, which has a
higher \ExtTSP score since all the edges are forward.

What is the computational complexity of Algorithm~\ref{algo:order}? A naive implementation
takes $\Oh(|V|^5)$ steps: There are $|V|$ merge iterations that process at most $|V|^2$ 
pairs of chains per iteration with $\Oh(|V|^2)$ steps needed to compute a merge gain 
between two chains. However, this is an overestimation, as we argue below.
First observe that the $\exttsp$ score between two chains, $c_1$ and $c_2$, can be positive
only if there is a branch between $c_1$ and $c_2$; thus, the number of candidate chain pairs
for merging is upper bounded by the total number of branches in the control flow graph, $|E|$.
The second observation is that one can memoize the results of \textsc{ComputeMergeGain}
function and re-use them throughout the computation. It is easy to see that the merge
gain depends only on a pair of chains; hence, if neither of the two chains is merged
at an iteration, then we do not need to recompute the gain for the pair at the next iteration.
Putting the two observations together, the running time of Algorithm~\ref{algo:order}
is bounded by $\Oh\big(\sum_{c} \size(c) \cdot \degree(c) \big)$, where the sum is
taken over all chains taking part in a merge with $\size(c)$ being the number of
blocks in the chain and $\degree(c)$ being the number of branches from and to the chain.
In the worst case, this sums up to $\Oh(|V|^2|E|)$ time in general, which equals to
$\Oh(|V|^3)$ for real-world control flow graphs.

\subsection{Large functions}
While cubic running time is acceptable for reordering most of the functions we
experimented with, there are several exceptions with a large number 
of basic blocks. In order to deal with these cases,
we introduce a threshold, $k > 1$, on the maximum size of a chain that is
considered for splitting in \textsc{ComputeMergeGain}. If the size
of a chain, $c$, exceeds the threshold, that is, $\size(c) > k$, then we only try a
simple concatenation of $c$ with other chains. With the modification, 
the complexity of Algorithm~\ref{algo:order} is estimated by $\Oh(k \cdot |V|^2)$, which
is quadratic when $k$ is a constant.
In our implementation, we use $k=128$ as the default value.

\subsection{Reordering of cold blocks}
Notice that Algorithm~\ref{algo:order} is not trying to optimize layout of cold basic blocks that
are never sampled during profiling. However, one may still want to modify their relative order, as
this could affect the code size as follows. Consider pairs of cold basic blocks, $s$ and $t$,
such that the only outgoing branch from $s$ is the only incoming branch to $t$. If $s$ and $t$ are
not consecutive in the resulting ordering, then one would need to introduce an unconditional branch
instruction. In contrast, if $t$ follows $s$ in the ordering, then the instruction
is not needed as $t$ is on the fall-through path of $s$. In order to guarantee that
Algorithm~\ref{algo:order} always merges such pairs into a chain, we modify 
the weights of cold edges in the control flow graph before the computation. Specifically we
set $w(s, t) = \epsilon_1$ (for some $0 < \epsilon_1 \ll 1$) if $(s, t) \in E$
corresponds to a cold fall-through branch in the original binary, and set
$w(s, t) = \epsilon_2$ (for some $0 < \epsilon_2 < \epsilon_1$) if $(s, t)$
corresponds to a cold non-fall-through branch. Such weights make it
desirable to merge original fall-through branches, even if they are cold according to the 
profile.

\subsection{Code layout in memory}
Apart from basic block reordering, profile-guided optimization tools typically perform
two other passes directly affecting the layout of functions in the generated code:
hot/cold code splitting and code alignment~\cite{DXT16,PANO18}. The first one
splits hot and cold basic blocks into separate sections, while the second pass
aligns the blocks at cache line boundaries via introducing NOP instructions.
We stress that both optimizations are complimentary to basic block reordering and
their benefits are additive. In our experiments, we evaluate the effect of reordering
alone, with all other optimization passes applied for the binary.

\section{An Optimal Algorithm for \ExtTSP}
\label{sect:mip}

We now demonstrate how \ExtTSP is formulated as a Mixed Integer Program (MIP).
A MIP is a method to find optimal solutions for \NP-hard problems whose objective and requirements are
represented by linear functions.
This is a time-consuming technique that can be applied only for small instances, 
and we use the optimal MIP solutions to better understand the quality of our heuristic.

$$
\begin{array}{lrcll}
\text{maximize} & \multicolumn{4}{l}{\sum_{(s, t)} w(s, t) \times f(d_{s, t})}\\
\text{subject to}&x_{s} & \in & \mathbb{R}, & s \in V \\
&d_{s,t} & \in & \mathbb{R}, & (s,t) \in E \\
&z_{s, t} & \in & \{0, 1\}, & s,t \in V, s \ne t\\
&x_t - x_s & \ge & L_s - M & \\
&&& \times (1 - z_{s, t}), & s,t \in V, s \ne t \\
&x_s - x_t & \ge & L_t - M \cdot z_{s, t}, & s,t \in V, s \ne t \\
&d_{s,t} & = & x_t - x_s - L_s, & (s,t) \in E \\
\end{array}
$$

The objective is a summation over the contribution of each edge $(s,t) \in E$ in the control flow graph to \ExtTSP.
The contribution of an edge is the number of jumps, $w(s,t)$, weighted by a value dependent on the length of the jump, 
$d_{s,t} = \len(s, t)$.
The piece-wise function shown in Fig.~\ref{fig:featuresb} converts the length to the desired weight.
This function is formulated in MIP by introducing additional integer variables to cope with the 
non-convex shape~\cite{CGM03}.

The constraints express the complete search space of all legal starting bytes $x_s$ for each block $s \in V$ considering the 
size of the block, $L_s$.
For all pairs of blocks, $s$ and $t$, either the final ordering has $s$ before $t$ (that is, $x_t - x_s \ge L_s$) or $t$ before~$s$ (that is, $x_s - x_t \ge L_t$).
A binary variable $z_{s,t}$ is utilized to enforce one of those two constraints.
The distances used in the objective, $d_{s,t}$, are constrained to be the distance between the end of block $s$ and the start of block $t$.
Negative distances correspond to backward jumps which are incorporated into the piece-wise function in the objective.
We utilize the Xpress solver for finding optimal solutions of the MIP model.

\section{Evaluation}
\label{sect:exp}

The experiments presented in this section were conducted on Linux-based servers powered by Intel
microprocessors. The applications were compiled using either GCC~8.3 or Clang~7.1 with -O3 optimization
level.

\subsection{Techniques}
\label{sect:techniques}
We compare our new algorithm (referred to as \alg{ext-tsp}) with the following competitors.

\begin{itemize}[leftmargin=6mm]
	\item \alg{original} is the ordering provided by the compiler.
	
	\item \alg{tsp} is the ordering constructed by the ``top-down'' heuristic suggested by Pettis
	and Hansen~\cite{PH90}. The algorithm starts by placing the entry basic block for a function,
	and then iteratively finding a successor with the heaviest edge to the last placed block. If all
	successors have already been selected, then one picks the block with the largest connection to the already
	placed blocks.
	
	\item \alg{ph} is the Pettis-Hansen ``bottom-up'' algorithm~\cite{PH90}. The algorithms maintains
	a collection of chains of basic blocks, which correspond to paths in the control flow graph.
	Initially every block forms its own chain. Looking at the arcs from largest to smallest, two different chains
	are merged together if the arc connects the tail of one chain to the head of another. Once the merging 
	stage is done, the chains are ordered so as to maximize the weight of
	backward edges to achieve the best performance of the branch predictor.
	
	\item \alg{cache} represents a modification of the Pettis-Hansen algorithm suggested by Luk et al. for the
	Ispike post-link optimizer~\cite{LMPCL04}. The difference from \alg{ph} is in the last step, ordering of
	chains of basic blocks. The chains are sorted by their density, that is, the total execution count
	of a chain divided by the sum of sizes of its basic blocks. Placing hottest chains first reduces 
	conflicts in the I-cache and improves code locality.
	
	\item \alg{mip} is an optimal algorithm for $\exttsp$ described in Section~\ref{sect:mip}. Since the
	running time of the approach is not practical for large functions, we only compare the results
	of \alg{mip} on a subset of small functions.
\end{itemize}

All the algorithms are implemented in an open-source post-link binary optimizer BOLT~\cite{BOLTSrc}.

\subsection{Facebook Workloads}

\begingroup
\setlength{\textfloatsep}{-2pt} 
\begin{table}[!t]
	\small
	\centering
	\caption{Basic properties of evaluated binaries}
	\label{table:dataset}
	\begin{tabular}{lrrrrrr}
		\toprule
		\centering 
		& & & \multicolumn{1}{c}{hot} & \multicolumn{3}{c}{blocks per function} \\
		& .text (MB) & IPC & functions & p50 & p95 & max \\
		\midrule
		\textsf{HHVM}	  & $285$  & $0.83$   & $12,\!687$ & $40$ & $454$ & $10,\!664$ \\
		\textsf{Multifeed} & $395$  & $0.95$   & $24,\!037$ & $15$ & $151$ & $9,\!228$ \\
		\textsf{Proxygen}	  & $160$  & $0.63$   & $9,\!997$ & $10$ & $97$ & $945$ \\
		\midrule
		\textsf{Clang}	  	  & $48$   & $0.82$   & $7,\!013$ & $25$ & $241$ & $11,\!218$ \\
		\textsf{GCC}	  	  & $15$   & $0.76$   & $10,\!269$ & $11$ & $152$ & $3,\!354$ \\
		\bottomrule
	\end{tabular}
\end{table}

\endgroup

This section evaluates various basic block ordering algorithms on four large-scale binaries
deployed at Facebook's data centers. The first system, 
which is our primary use case for \alg{ext-tsp}, is the
HipHop Virtual Machine (\textsf{HHVM})~\cite{AEM14}, that serves as an execution engine for PHP
at Facebook, Wikipedia, Baidu, and other large websites. The two binaries of
\textsf{Multifeed} are responsible for News Feed. \textsf{Proxygen} is a Facebook service for
cluster load balancing.
The \textsf{HHVM} binary is built using GCC with LTO, while \textsf{Multifeed} and \textsf{Proxygen}
are compiled with Clang with AutoFDO enabled to enhance their performance.
All the Facebook services are running with huge pages enabled and utilize function reordering~\cite{OM17}.
Table~\ref{table:dataset} provides basic properties of the evaluated binaries.

\begin{figure}[t]
	\centering	
	\includegraphics[width=0.85\columnwidth]{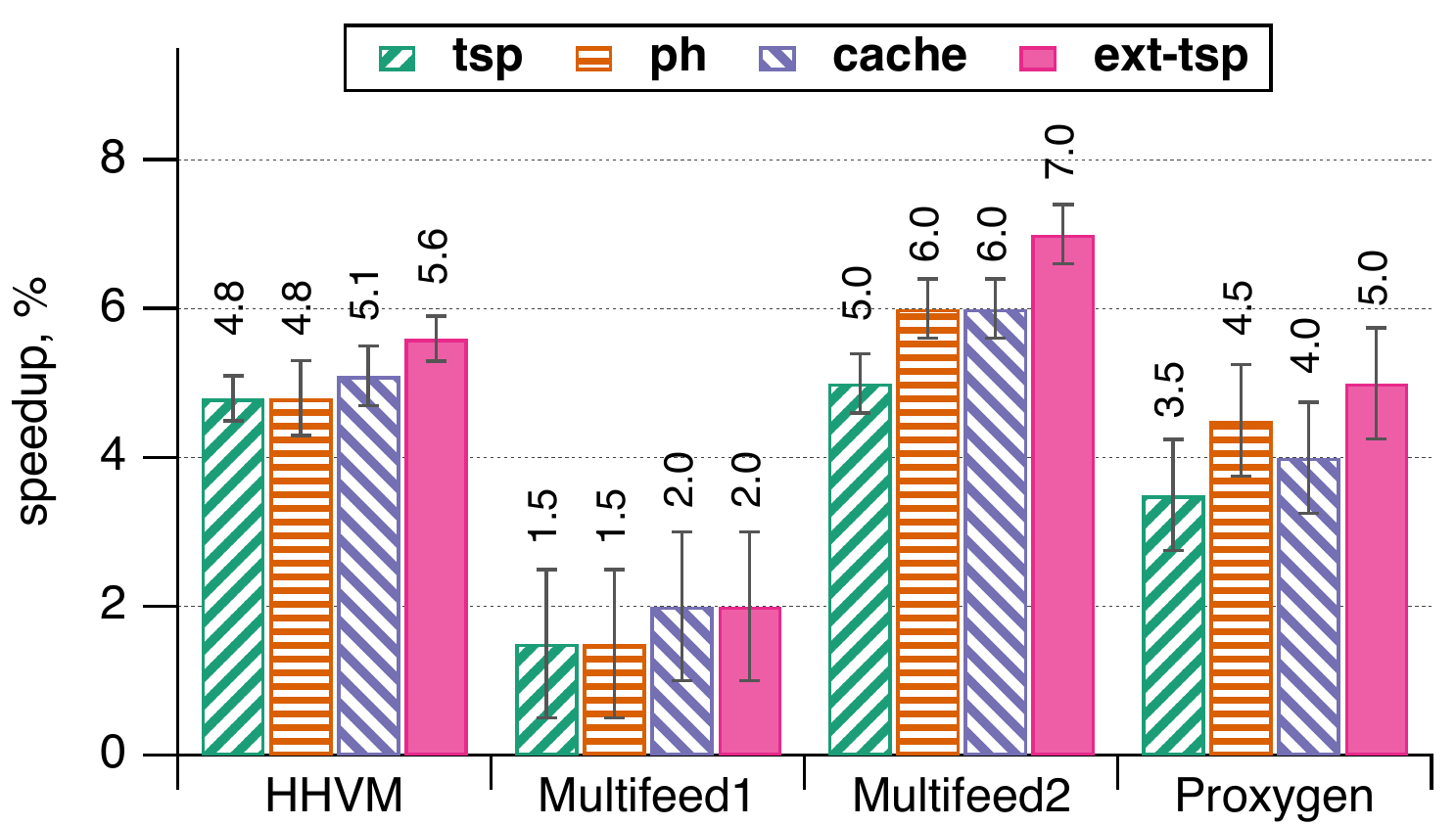}
	\caption{Performance improvements of various reordering algorithms over \alg{original} measured for different Facebook workloads.}
	\label{fig:improvementsb}
\end{figure}

Fig.~\ref{fig:improvementsb} presents a performance comparison of four basic block ordering
algorithms on the Facebook workloads. The results are obtained by using an internal performance-measurement
tool for running A/B experiments; see~\cite{BF15,LKOB18} for an overview.
The tool is used at Facebook for a wide range of performance evaluations by running experiments on a set
of isolated machines that process the same production traffic over several days. We measure
performance as the CPU utilization during steady state.
As a baseline, we observe the performance of the binaries optimized with BOLT using the 
\alg{original} block ordering algorithm. In the case of \textsf{HHVM}, this is
an original ordering constructed by the compiler, while for \textsf{Multifeed} and \textsf{Proxygen}, 
the ordering is a result of processing the binaries with PGO.
The figure reports mean relative improvements of various block ordering algorithms over \alg{original}
along with their $95\%$ confidence intervals.
In the experiments, we notice that differences in CPU utilization among the block ordering algorithms are 
highly correlated with the differences in instructions per cycle (IPC).

\begin{figure}[t]
	\centering	
	\includegraphics[width=0.95\columnwidth]{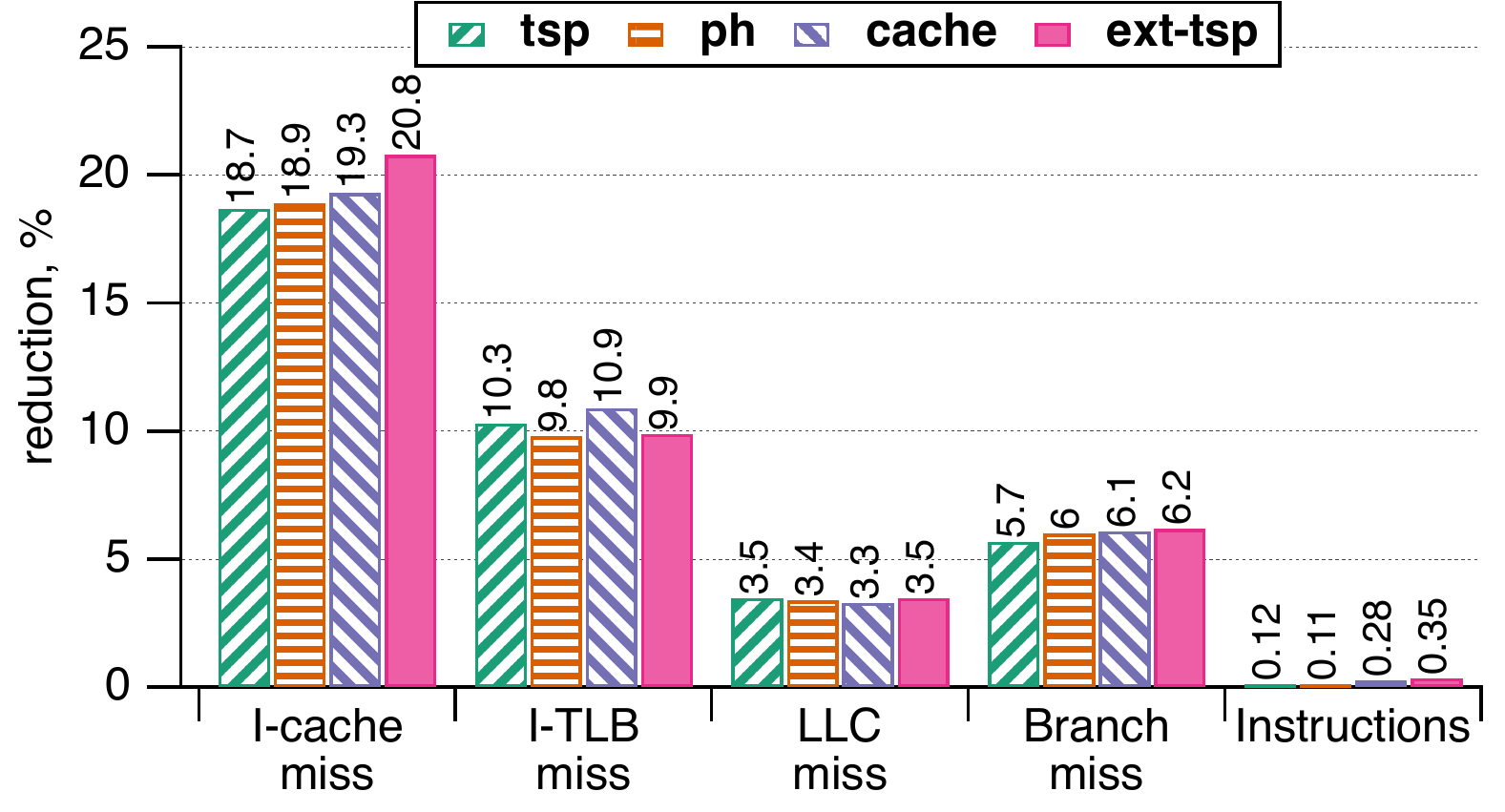}
	\caption{\texttt{perf} metrics measured for the \textsf{HHVM} binary.}
	\label{fig:improvementsa}
\end{figure}

Overall we observe that \alg{ext-tsp} performs better than alternative ordering algorithms on three of the
evaluated binaries. The relative speedup is close to $1\%$ for \textsf{Multifeed2} and 
around $0.5\%$ for \textsf{HHVM} and \textsf{Proxygen}. We stress that the measurements for 
\textsf{Multifeed1} are noticeably noisier than the alternatives with a typical deviation from the mean around
$0.5\%{-}1\%$. To better understand the benefits of applying the new block ordering algorithm,
we perform a more detailed evaluation of \textsf{HHVM}. The results are 
depicted in Fig.~\ref{fig:improvementsa}.
The main advantage of block ordering optimization is an improved performance of the L1 I-cache, 
that exhibits over $19\%$ miss reduction. The new ordering algorithm increases this value
to $21\%$.
The number of branch and I-TLB misses is also significantly reduced, with \alg{ext-tsp}
being the best for the branch misses counter. We also see a modest improvement in
the performance of the last level cache, though the difference between
various ordering algorithms is not prominent.

\subsection{Open-Source Compilers}
Since basic block reordering primarily improves the performance of the I-cache, 
our optimization can be beneficial for any front-end bound application with large code size.
We illustrate this by experimenting with binaries of two open-source compilers, Clang and GCC, 
whose \texttt{.text} sections are $48$MB and $15$MB, respectively; see Table~\ref{table:dataset}.
For these experiments, we utilize a dual-node 28-core 2.4 GHz Intel Xeon E5-2680 (Broadwell)
with $256$GB RAM. The size of the L1 I-cache on the processor is only $32$KB, which
makes the two binaries good candidates for profile-guided layout optimizations.

\begin{figure}[t]
	\centering	
	\includegraphics[width=0.8\columnwidth]{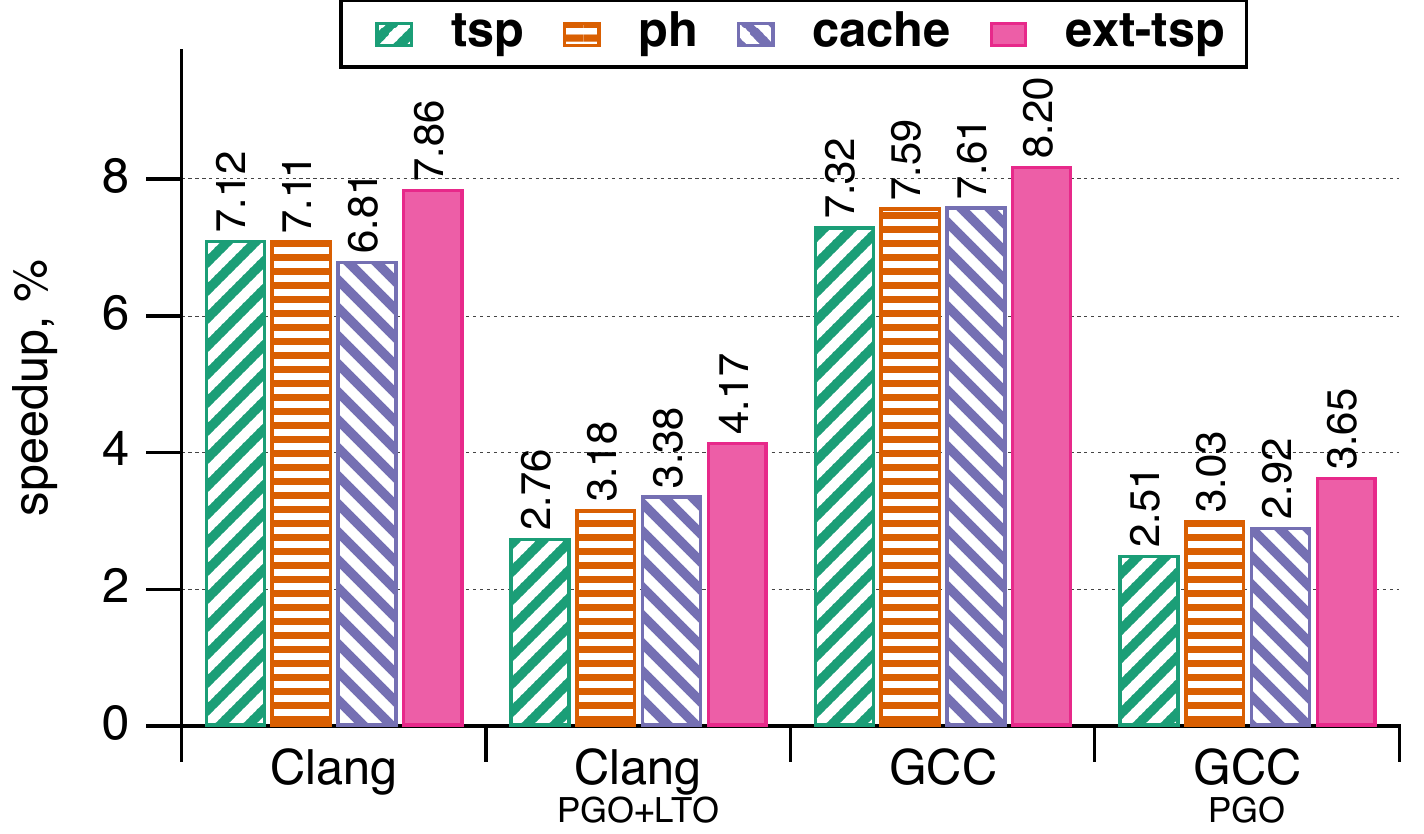}
	\caption{Performance improvements of various reordering algorithms over \alg{original} 
		measured for \textsf{Clang} and \textsf{GCC}.}
	\label{fig:impr_compilers}
\end{figure}

For the evaluation of Clang, we use the \texttt{release\_71} branch of LLVM.
We ran the evaluation in two modes, \textsf{Clang} and \textsf{Clang+PGO+LTO};
see Fig.~\ref{fig:impr_compilers}. First we build a release version of the binary using GCC
and collect a profile data by compiling a medium-sized template-heavy C++14 source file.
For the first experiment (\textsf{Clang}), we optimize the binary of Clang with BOLT
using various orderings of basic blocks, and compile a different collection of about $100$ 
C++14 source files. Hence, the train and the test datasets are different in the evaluation.
Every experimental run is repeated $1000$ times to increase precision of our measurements so 
that the average mean deviation is within $0.05\%$.
The baseline in Fig.~\ref{fig:impr_compilers} 
corresponds to a binary processed with BOLT using the \alg{original} ordering of basic blocks.
Thus, the improvements are attributed only to block reordering, while all other optimizations
(e.g., function reordering or inlining) are the same across the experiments.

\begin{figure}[t]
	\centering
	\subfloat{
		\includegraphics[width=0.95\columnwidth]{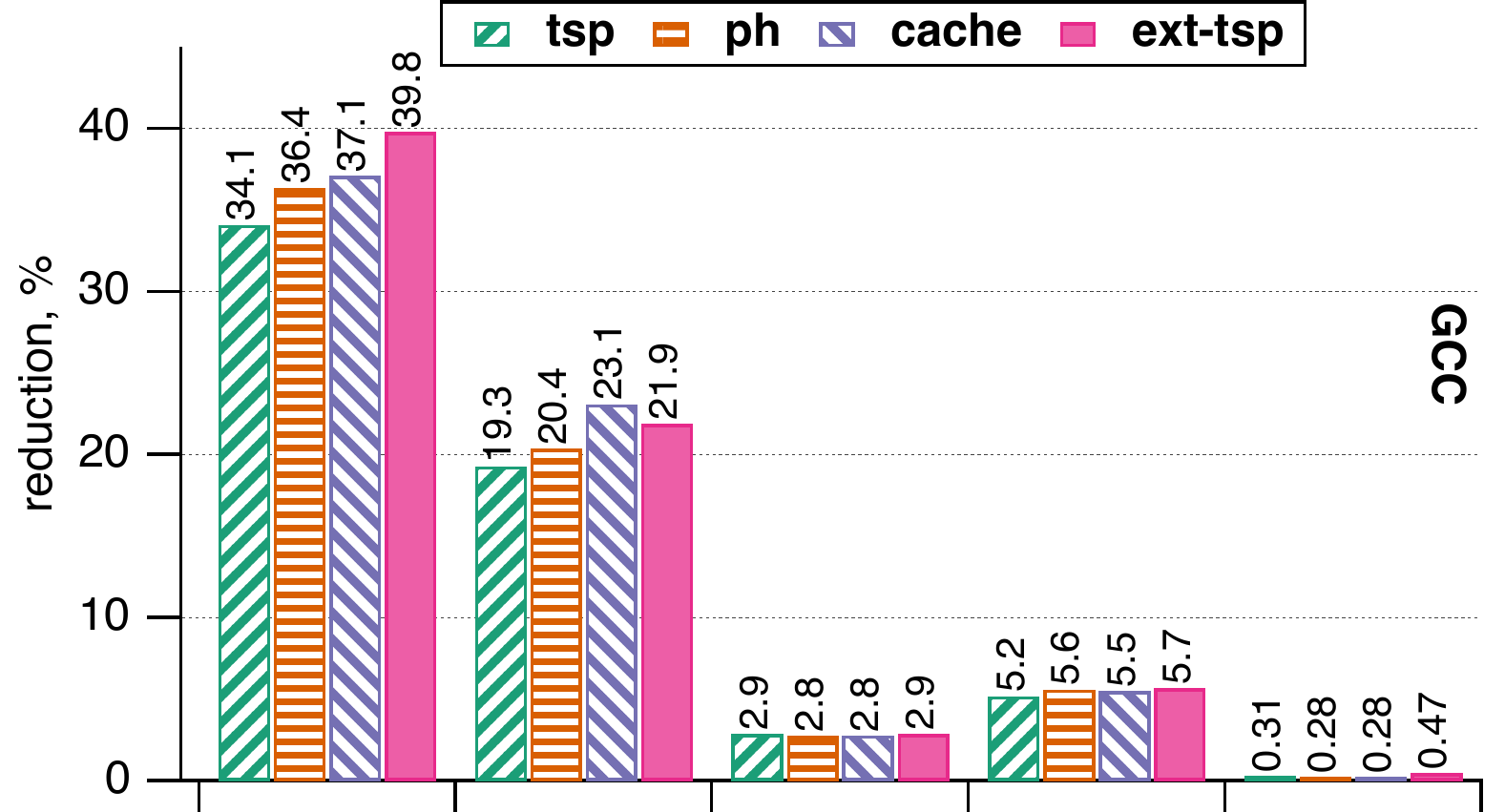}
	}
	\hfill
	\subfloat{
		\includegraphics[width=0.95\columnwidth]{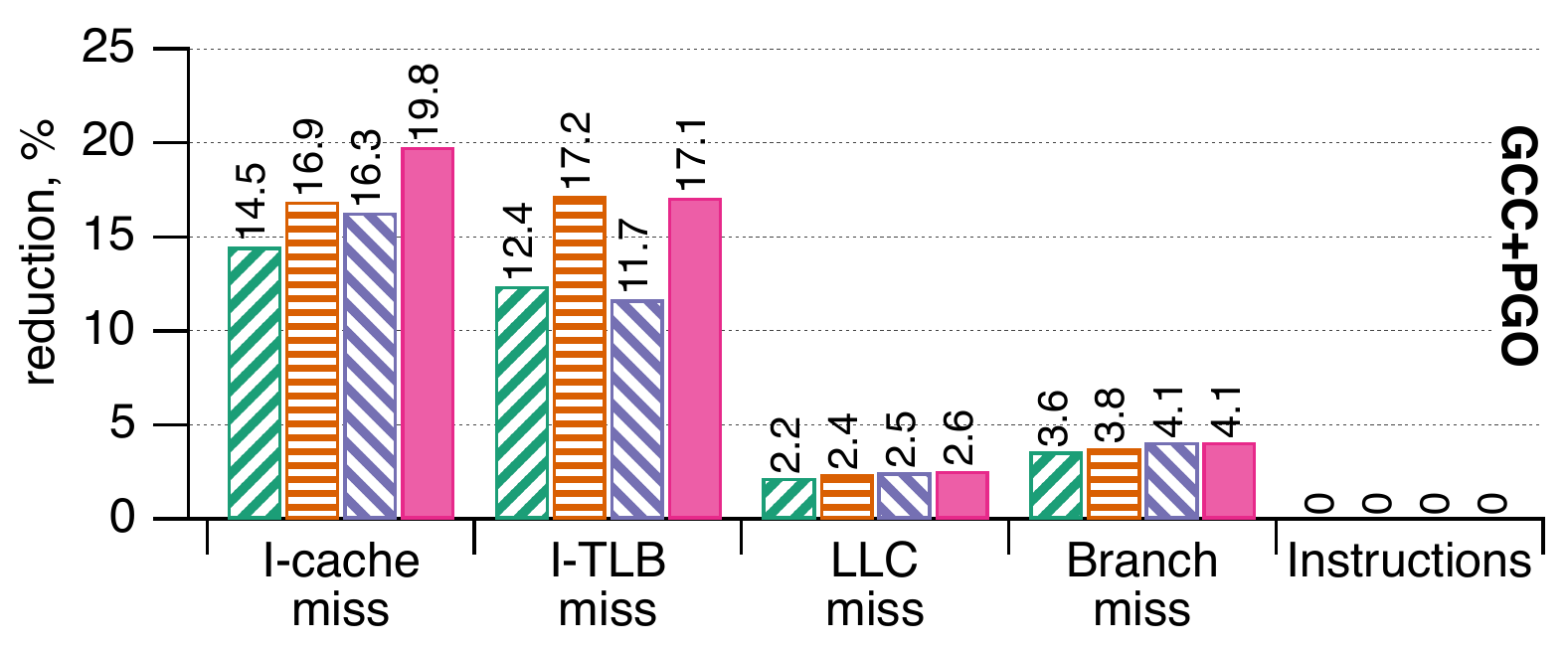}
	}	
	\caption{\texttt{perf} metrics measured for the \textsf{GCC} binary
		with (bottom) and without (top) \textsf{PGO} applied. The improvements are
		on top of the \alg{original} block ordering using the same configuration.}
	\label{fig:improvementsa3}
\end{figure}

For our next experiment, \textsf{Clang+PGO+LTO}, we in addition enable PGO and LTO support.
To this end, we first built an instrumented version of Clang, and then used the instrumented 
compiler to build the binary again with the default options of GCC. The collected profile data 
was used to do another build of Clang with PGO and LTO enabled.
The results in Fig.~\ref{fig:impr_compilers} 
indicate that block reordering alone provides $2\%-4\%$ performance improvements, even when applied on top of
GCC with PGO and LTO. This finding is consistent with earlier evaluations of BOLT, where
the gains are attributed to an improved code layout~\cite{PANO18}.

For the evaluation of GCC, we use version \texttt{8.3} of the compiler. Again, we collected a profile data
by compiling the single source file, and perform testing using the larger C++14 project.
Similar to Clang, we ran experiments in two modes; however, we did not use LTO due to build errors
related to C++ exceptions.
The effect of applying various block ordering techniques on two versions of the GCC binary
(with and without PGO enabled) is shown in Fig.~\ref{fig:impr_compilers}. As expected,
the relative improvements are smaller for the binary built with PGO. The new algorithm, \alg{ext-tsp}, 
provides the largest gains outperforming competitors by $0.5\%-0.8\%$; the differences are identified as
statistically significant. Fig.~\ref{fig:improvementsa3} presents the impact of block ordering
algorithms on key architectural metrics for GCC.
As in the case with \textsf{HHVM}, the improvements are largely attributed to a reduction in 
I-cache and I-TLB misses. Other relevant metrics are also improved in comparison with the
\alg{original} ordering but there is little difference among the ordering techniques.

\subsection{SPEC CPU 2017}

In this section we evaluate basic block reordering on the SPEC CPU 2017 benchmark. 
We utilize $16$ C/C++ programs compiled using GCC with LTO and ran experiments on the same hardware
as in the previous section. We analyze the performance of the binaries optimized by
BOLT with various ordering algorithms, using \alg{original} as a baseline.
Profile data is collected using a separate SPEC train mode.

\begin{figure}[t]
	\includegraphics[width=1.0\columnwidth]{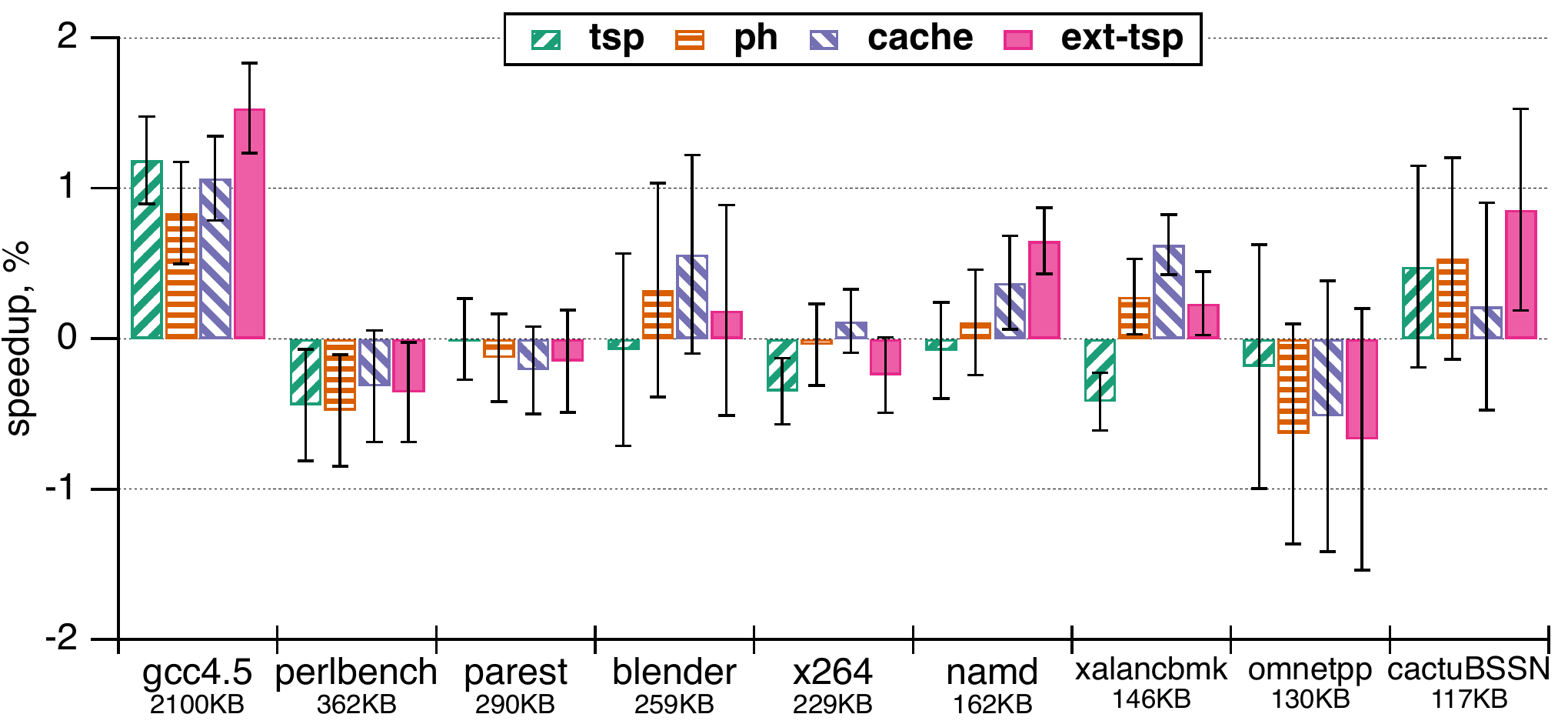}
	\caption{Relative performance differences between \alg{original} and alternative block reordering methods on the largest
		binaries of the SPEC 2017 dataset. Positive values indicate improvements, negative ones indicate regressions.
		For every binary, the size of the hot code is specified.}
	\label{fig:spec}
\end{figure}

We observe that the SPEC binaries are much smaller than the typical applications used in modern 
data centers. Therefore, they are unlikely to be front-end bound and exhibit many I-cache and I-TLB misses. 
Fig.~\ref{fig:spec} presents the results of our experiments on the largest binaries 
that contain at least $100$KB of hot code according to the collected profile.
We do not see a consistent advantage of applying basic block reordering for the binaries. In most of the
experiments we record a high variance in the running times, which
often exceeds the differences between means. An optimized block ordering yields a statistically
significant improvement over \alg{original} only for three binaries: \texttt{gcc4.5} and \texttt{namd} (using
\alg{ext-tsp}), and \texttt{xalancbmk} (using \alg{cache}).
For the largest program, \texttt{gcc4.5}, the \alg{ext-tsp} algorithm achieves $1.5\%$ speedup outperforming the 
best competitor by $0.4\%$.

To understand the source of regressions, we analyze two binaries from the benchmark, \texttt{x264} and \texttt{omnetpp},
whose running times increase by $0.5\%-1.5\%$ after block reordering. In the former case, we observe a
substantial growth in the number of branch misses, which leads to the performance regression. In the latter case, 
we found that a different alignment of hot loops in the binary worsen the performance despite a
significant increase ($20\%$) of the number of fall-through branches and a modest improvement (around $5\%$)
in the number of I-cache misses.
We conclude that for small binaries that are not front-end bound, both \TSP and \ExtTSP are not accurate models.


\subsection{Analysis of \ExtTSP}

Here we present an evaluation of Algorithm~\ref{algo:order} for solving the \ExtTSP problem.
We design the experiments to answer two questions:
(a)~How do various parameters of the algorithm contribute to the solution and what are the best default values?
(b)~How does the algorithm perform in comparison with existing heuristics and the optimal technique?

Considering the first question, we observe that \alg{ext-tsp} has only one parameter that affects 
its quality and performance. As explained in Section~\ref{sect:algo}, 
we introduce a threshold $k$, which controls the maximum size of a chain that can be split during
optimization. In the extreme case with $k=|V|$, all chains can be broken if that improves the objective;
however, the running time of the algorithm is cubic on the number of basic blocks comprising a function.
Another extreme, $k=0$, forbids chain splitting but makes the running time quadratic. As Table~\ref{table:dataset}
illustrates, some functions in the dataset contain a few thousand of basic blocks. Hence, the threshold should
be chosen carefully, since it impacts the quality of a solution and the time needed to process a binary, which
is important in production environments.

\begin{figure}[t]
	\centering	
	\subfloat{
		\includegraphics[width=0.5\columnwidth]{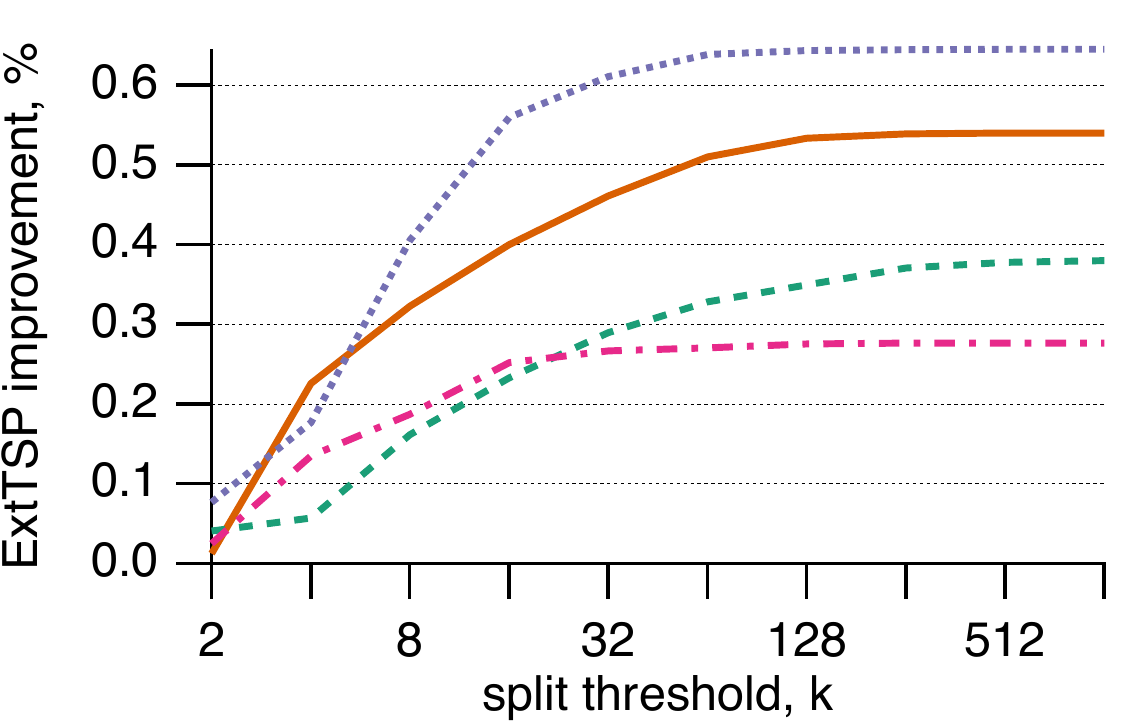}
	}
	\subfloat{
		\includegraphics[width=0.5\columnwidth]{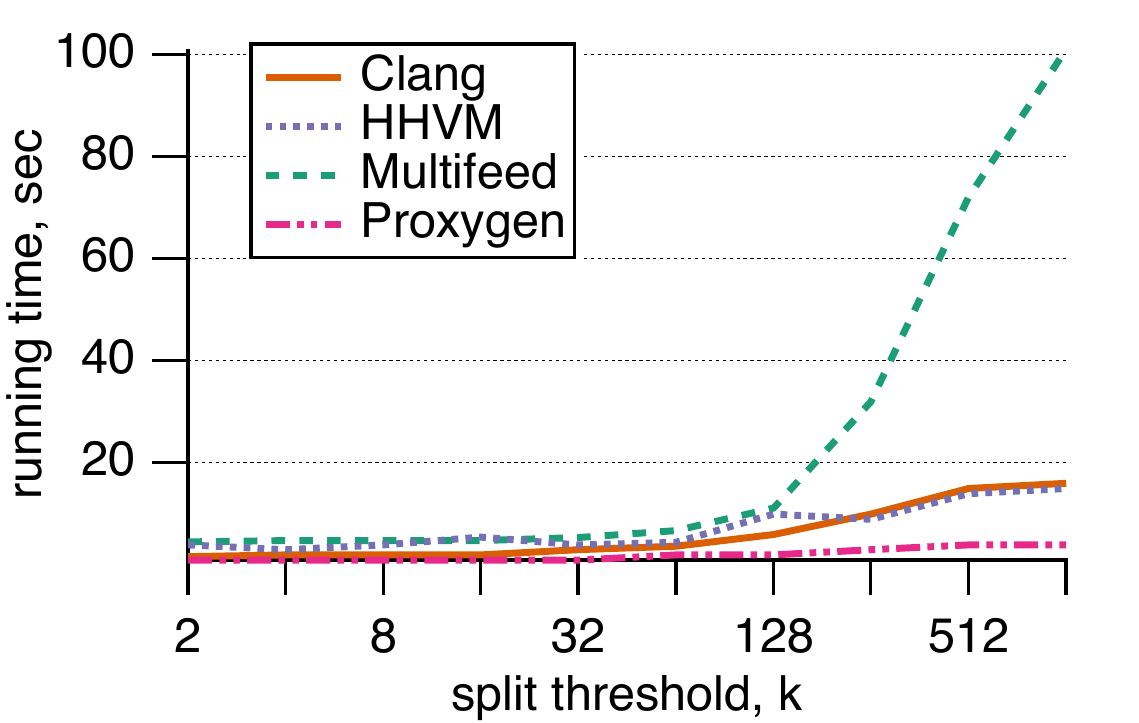}
	}
	\caption{The running times and the resulting \ExtTSP scores produced by \alg{ext-tsp} (Algorithm~\ref{algo:order})
		using various chain splitting thresholds for binaries described in Table~\ref{table:dataset}.}
	\label{fig:threshold}
\end{figure}

Fig.~\ref{fig:threshold} illustrates the results of the experiments with the chain splitting threshold.
\textsf{Multifeed1} is the binary whose processing time is substantially affected by large values of~$k$.
For $k \ge 1024$, the combined running time of \alg{ext-tsp} on all functions of the \textsf{Multifeed1} binary is
around $2$ minutes, while for \textsf{HHVM} it is less than $20$ seconds. For $k=0$, the processing times
are $6$ and $4$ seconds for the binaries, respectively.
The difference between the corresponding 
solutions in the \ExtTSP score ranges between $0.3\%$ and $0.7\%$; note that according to our
analysis in Section~\ref{sect:ml} and Fig.~\ref{fig:cpu_vs_tsp}, this translates to a performance
difference of $0.1\%-0.3\%$. 
The value $k=128$ provides a reasonable compromise between processing speed and solution quality, and thus, 
it is utilized as the default value for \alg{ext-tsp} in all our experiments.


In order to analyze the quality of solutions for \ExtTSP generated by Algorithm~\ref{algo:order}, we
employ the optimal technique presented in Section~\ref{sect:mip}.
We apply \alg{mip} to all $2992$ functions containing at most $30$ basic blocks in the \textsf{HHVM} binary.
For these small functions, \alg{mip} finds a provably optimal solution in $2963$ ($99\%$)
of the cases in less than one minute.
Out of these instances with the known optimal ordering, \alg{ext-tsp} finds an
equivalent solution in $2914$ ($98.3\%$) cases. For the remaining $49$ functions, the \ExtTSP score
produced by \alg{ext-tsp} is on average $0.14\%$ lower than the optimum.
For comparison, the runner-up approach on the same binary is \alg{cache}, which is able to
reconstruct $2745$ ($92.6\%$) of optimal orderings in the binary.
The relative improvements in the \ExtTSP score over non-reordered functions are $27\%$, $29\%$, $29\%$, $31\%$
for \alg{tsp}, \alg{ph}, \alg{cache}, and \alg{ext-tsp}, respectively, which
aligns with our experiments illustrated in Fig.~\ref{fig:improvementsb}.

We emphasize that \alg{mip} is not considered a practical approach, as it does not scale
to instances with many basic blocks. The average running time of
\alg{mip} on a function with $30$ blocks exceeds $10$ seconds, while it is below a millisecond
for all four alternatives, \alg{tsp}, \alg{ph}, \alg{cache}, and \alg{ext-tsp}.
Nevertheless, the aforementioned analysis demonstrates that the new heuristic provides
a close-to-optimal solutions in the majority of real-world instances, while
being sufficiently fast to process large functions.

\section{Related Work}
\label{sect:related}

There exists a rich literature on profile-guided optimizations. Here we discuss previous works that 
are closely related to code layout and our main contributions.

The work by Pettis and Hansen~\cite{PH90} is the basis for the majority of modern code
reordering techniques. The goal is to create chains of basic blocks that are frequently executed 
together in the order. As discussed in Section~\ref{sect:ml}, many variants of the technique
have been suggested in the literature and implemented in various tools~\cite{BG77,YJSK97,HMK79,CG94,TXD98,LMPCL04,SDAL01,RLN14}.
Similar to our work, the techniques are operating with a control flow graph and try to lay out
basic blocks tackling a variant of the \prob{Traveling Salesman Problem}. Alternative models
have been studied by Bahar et al.~\cite{BCG99}, Gloy et~al.~\cite{GS99}, and Lavaee and Ding~\cite{LD14}, 
where a temporal-relation graph is taken into account. Temporal affinities between code instructions can be 
utilized for reducing conflict cache misses~\cite{HKC97} and improving the performance
of multiple applications using a shared cache~\cite{LLDHY14}. We emphasize that according to our
experiments, the performance of a front-end bound large-scale binary can be 
largely predicted by its control flow graph without considering more expensive models.

Code reordering at the function-level is also initiated by Pettis and Hansen~\cite{PH90}, who
describe an algorithm that is implemented in many compilers and binary optimization
tools~\cite{RLN14,PANO18,RBG01,SRM98}. This approach greedily merges chains of functions and
is designed to primarily reduce I-TLB misses. An improvement is recently proposed by
Ottoni and Maher~\cite{OM17}, who suggest to work with a directed call graph. 
Note that unlike our work, the techniques are heuristics not
aiming to produce code layouts that are optimal from the performance point of view.

Another opportunity for improving performance is to modify layout of data~\cite{RHM07,EKMT16,RL16,CS06}.
Most of the existing works focus on field reordering and structure splitting based on the
field hotness and data affinities. While the problem of finding an optimal data layout is
computationally hard~\cite{PR02,Lav16}, we believe that utilizing machine learning
may lead to improved heuristics resulting in performance gains for real-world applications.

\section{Conclusion}
\label{sect:conclude}

In this work we extended the state-of-the-art model for reordering of basic blocks and developed a new
efficient algorithm to optimize the layout of a binary. We also performed an extensive evaluation of various
block ordering techniques on a variety of real-world applications. The experiments indicate that the
new technique can improve the performance of binaries that have been manually tuned over the course of
their development and optimized using conventional compiler optimizations.
There are several interesting aspects of our approach that we discuss next. 

Firstly, our approach employs a machine learning
toolkit to build a desired objective for optimization. As our evaluation demonstrates, 
the resulting model outperforms the classical one, as the new objective correlates very well with the
performance of large-scale binaries. A possible risk here is to over-tune a model for a specific application
and miss important details that might affect performance. The experiments with the SPEC benchmark 
imply that the models based on maximizing the number of fall-through branches are too simplistic
for binaries that are not front-end bound.

Secondly, our study focuses on optimizing applications built with particular compilers and
running on a specific hardware. A reasonable future work is to verify whether the presented approach
can be generalized to other use cases. Our preliminary experiments indicate that comparable gains can
be achieved on other Intel microprocessors and alternative processor architectures.
Similarly, the new reordering algorithm is applied as a post-link optimization, and we did not
examine how it behaves on earlier compilation stages. It would be interesting to investigate the effect
of the reordering applied at compilation time. In particular, we plan in the future to integrate 
and compare \alg{ext-tsp} with the algorithms implemented in GCC~\cite{TXD98} and Clang.

Finally, we point out that this paper considers a certain aspect related to code generation:
reordering of basic blocks within a function. There are many complementary optimizations
that we did not investigate in detail, for example, un-rolling loops or duplicating blocks in order
to avoid extra jumps. An attractive direction is to allow cross-procedure reordering
in which basic blocks from different functions can be interleaved in the final layout. This might
further increase code locality and improve cache utilization.
Unfortunately, our preliminary experiments with existing cross-procedure heuristics~\cite{TXD98} did not
produce measurable gains; further research of the technique is an intriguing future work.

\ifCLASSOPTIONcompsoc
  \section*{Acknowledgments}
\else
  \section*{Acknowledgment}
\fi

We thank Alon Shalita for fruitful initial discussions of the project.
We would also like to thank Rafael Auler and Maksim Panchenko for their help
with integrating the new technique into BOLT.

\ifCLASSOPTIONcaptionsoff
  \newpage
\fi

\IEEEtriggeratref{19}


\bibliographystyle{IEEEtran}
\bibliography{IEEEabrv,main-arxiv}

\end{document}